\documentclass[aps,prl,twocolumn,groupedaddress,showpacs,floatfix,superscriptaddress,longbibliography]{revtex4-2}
\usepackage[plainpages=false,pdfpagelabels,colorlinks=true,linkcolor=red,urlcolor=blue,citecolor=blue,pdftitle={Title},pdfauthor={},pdfdisplaydoctitle=true,pdfduplex=DuplexFlipLongEdge]{hyperref}
\usepackage{epsfig}
\usepackage{amsmath,amssymb}
\usepackage{physics}
\usepackage{graphicx}
\usepackage[dvipsnames,usenames]{color}
\usepackage[normalem]{ulem}
\usepackage{soul} % to use \st for strikeout
\usepackage{float}
\usepackage{diagbox}

\renewcommand\thesection{\arabic{section}}
\newcommand{\tcr}{\textcolor{red}}

\newcommand{\beginsupplement}{%
        \setcounter{table}{0}
        \renewcommand{\thetable}{S\arabic{table}}%
        \setcounter{figure}{0}
        \renewcommand{\thefigure}{S\arabic{figure}}%
}

\begin{document}
\title{Stripes and the Emergence of Charge $\pi$-phase Shifts in Isotropically Paired Systems}

\author{Jianhao Sun}
\affiliation{Beijing Computational Science Research Center, Beijing 100193, China}
\author{Tao Ying}
\email{taoying86@hit.edu.cn}
\affiliation{School of Physics, Harbin Institute of Technology, Harbin 150001, China}
\author{Richard T.~Scalettar}
\email{scalettar@physics.ucdavis.edu}
\affiliation{Department of Physics and Astronomy, University of California, Davis, CA 95616, USA}
\author{Rubem Mondaini}
\email{rmondaini@csrc.ac.cn}
\affiliation{Beijing Computational Science Research Center, Beijing 100193, China}

\begin{abstract}
The interplay of spin and motional degrees of freedom forms a key element in explaining stripe formation accompanied by sublattice reversal of local antiferromagnetic ordering in interacting fermionic models. A long-standing question aims to relate pairing to stripe formation, intending to discern the applicability of simple models that observe this phenomenon in understanding cuprate physics. By departing from fermionic statistics, we show that the formation of stripes is rather generic, allowing one to unveil its competition with superfluid behavior. To that end, we use a combination of numerical methods to solve a model of interacting hardcore bosons in ladder geometries, finding that once stripes are formed, either via external pinning or spontaneously, a sublattice reversal ($\pi$-phase shift) of \textit{charge} ordering occurs, suppressing the superfluid weight. Lastly, we show that when the Cooper pairs are not local, as in the attractive Hubbard model with finite interactions, auxiliary-field quantum Monte Carlo calculations show evidence of fluctuating stripes, but these are seen to coexist with superfluidity. Our results corroborate the picture that static stripes cannot be reconciled with pairing, unlike the case of fluctuating ones.
\end{abstract}

\maketitle
\paragraph{Introduction.--} Clarifying whether simplified lattice models capture the salient features of certain classes of high-temperature superconductors, such as the cuprates, has been at the forefront of scientific research in condensed matter physics for over three decades~\cite{Scalapino1986, Emery1987, Zhang1988}. One of the aspects that complicates this quest is the absence of controlled analytical methods in dimensions larger than one that can tackle the solution of the corresponding Hamiltonians. Additionally, such a problem is even more elusive because of the small energy scales separating competing orders, creating challenges for numerical simulations.

While much remains to be settled, a recurring feature of existing calculations, which reproduces experimental observations~\cite{Tranquada2020}, is the presence of charge stripes, wherein the doped holes unidirectionally condense over periodic regions in the lattice~\cite{Zaanen1989, Poilblanc1989, Machida1989, Kato1990, White1998a, White1998b, Chang2010, Zheng2016, Boxiao2017, Huang2017, Huang2018, Vanhala2018, Jiang2019, Mai2022, Xu2022, Schlomer2022, Mai2023, Xu2023, Jiang2024}. An additional aspect revealed by experimental and theoretical results is the emergence of a reversal of the sublattice magnetization across a stripe region in certain regimes of parameters, dubbed a $\pi$-phase shift (spin stripe). The common lore is that the high-temperature superconductors are characterized by intertwined orders~\cite{Fradkin2015}, whose fate of competition/cooperation in suppressing/aiding the pairing properties is a question yet to be definitively answered.

Recently, however, large-scale numerical studies that combine constrained path quantum Monte Carlo~\cite{Zhang1995} and density matrix renormalization group (DMRG) methods~\cite{White1992, White1993} to investigate the doped Hubbard model in large cylinders have started to converge toward the solution of this problem at the ground state~\cite{Boxiao2017, Qin2020, Xu2023}. Other approaches, such as unbiased auxiliary-field quantum Monte Carlo~\cite{Huang2017, Huang2018, Liu2021} or dynamical cluster approximations~\cite{Mai2022, Mai2023}, are limited to low-but-finite temperatures, owing to the emergence of the sign problem~\cite{Loh1990, Mondaini2021}. Finite-temperature extensions of the DMRG method (minimally entangled typical thermal states~\cite{White2009, Stoudenmire2010}), however, allow one to bridge this gap, corroborating the emergence of charge and spin stripes over different temperature ranges~\cite{Wietek2021}.

In extensions to strongly-coupled regimes, where a description in terms of a $t$-$J$ model is appropriate~\cite{Zhang1988}, the formation of stripes is also observed. Still, the pairing in the hole-doped regime seems unattainable~\cite{White1999, Jiang2021}, defying the expectation that this particular model has ``the right stuff''~\cite{Scalapino94} to describe the cuprate physics. Note, however, that the inclusion of next-nearest-neighbor terms for both hoppings \textit{and} exchange can change this picture, making superconductivity more robust in the hole-doped regime~\cite{Gong2021, HCJiang2021, HCJiang2023}. More recently, it has also been noted that observing phase reversal across hole-rich regions is not unique to repulsive models but is too seen on the charge degrees of freedom instead on the \textit{attractive} Hubbard model~\cite{Ying2022}. 
%% In such case, similarly heading towards 
Here, in the strong-coupling limit, the typical size of the Cooper pairs becomes increasingly small, asymptotically approaching a regime where a description of repulsive hardcore bosons (a composite fermionic pair) is relevant~\cite{Micnas1990}. 
\begin{figure*}[t!]
    \centering
    \includegraphics[width=2.0\columnwidth]{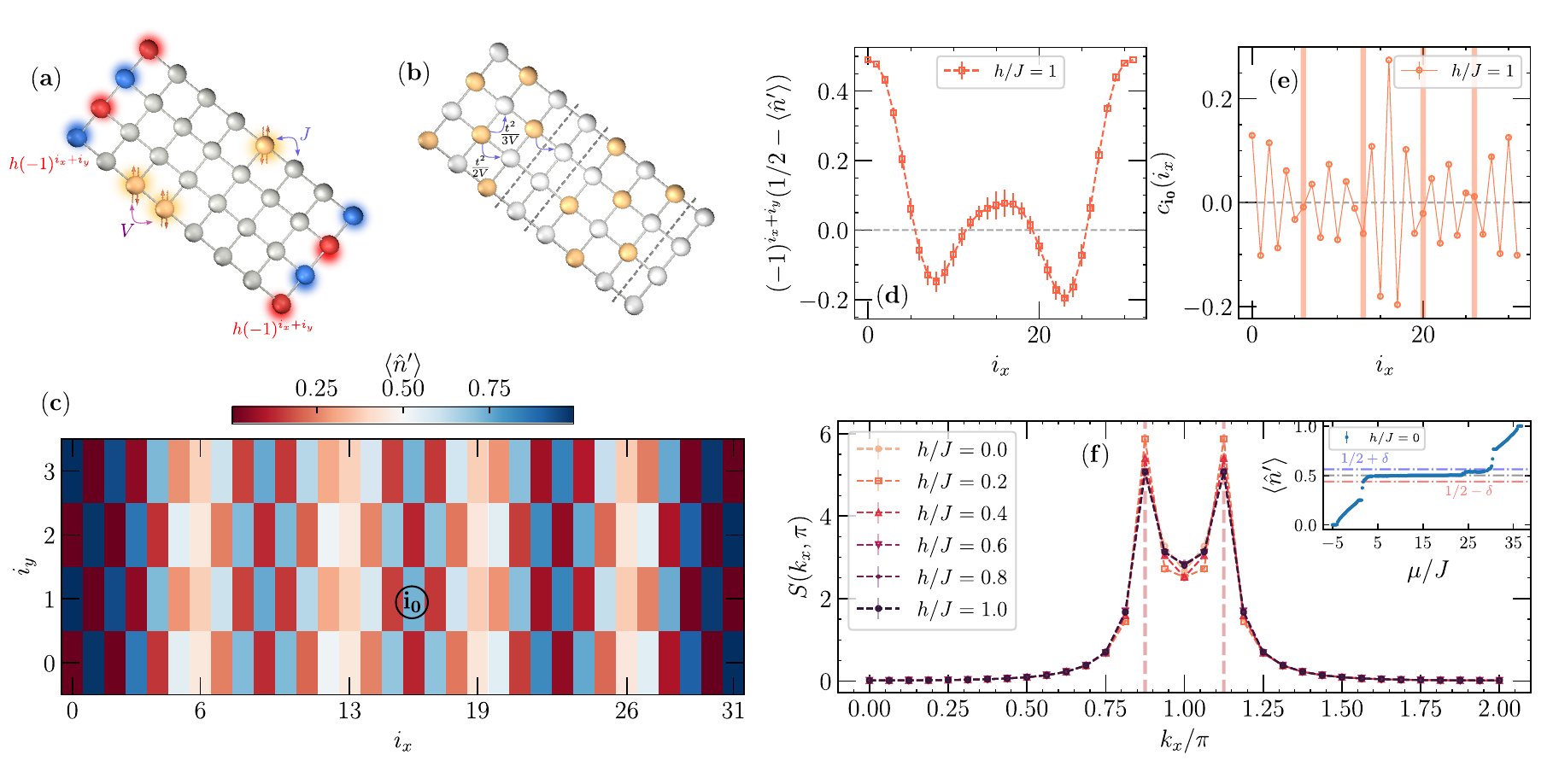}
    \caption{
    (a) Schematic representation of the lattice with relevant terms of the hardcore boson Hamiltonian annotated, including staggered pinning potential at the edges with magnitude $h$. (b) Representation of the `charge $\pi$-phase shift' that emerges in the strongly interacting limit ($V/J\gg1$), with contributions in second-order perturbation theory indicated. While the kinetic energy gains within a checkerboard domain are $\propto t^2/nV$ ($n=2,3$) for a given site, bosons can freely hop across the dashed lines marking the stripe locations. (c) density profile in $32\times 4$ ladder exhibiting four hole-regions in its short-direction; here $h/J = 1$. (d) The average staggered hole-density profile and (e), the corresponding density-density correlation functions with respect to site ${\bf i}_0$ annotated in (c), along $i_y=1$. (f) The density structure factor $S(k_x,\pi)$ shows robust peaks at $k_x=\pi (1\pm2\delta)$ even in the absence of pinning potential ($h=0$); the inset shows the phase separation characteristics for the studied densities emerging in the equation of state ($\langle n^\prime\rangle$ vs.~$\mu$). All data are obtained for $V/J=8$.}
    \label{fig:fig_1}
\end{figure*}

Many of the limitations that prevent the numerical study of the interplay of stripes and pairing do not occur if considering the attractive Hubbard Hamiltonian~\footnote{So long as the single-particle part is the same for both spin species.} or repulsive hardcore bosons without frustration. Furthermore, from an experimental point of view, the study of the competition between pairing and stripes/charge ordering is friendlier in the scope of quantum emulators based on ultracold atoms in optical lattices since the temperature scales for the onset of isotropic ($s$-wave) superfluidity are higher, potentially allowing its observation within existing regimes accessible~\cite{Hartke2022}, even if not directly aiming to tackle the physics of the cuprate-like materials. In this paper, we take advantage of the capabilities of quantum simulations of such models; our key conclusions are: (i)  In the strong coupling (bosonic) limit, {\it static} stripe density patterns emerge with doping; (ii) the charge density wave correlations are characterized by a phase shift across the stripes for sufficiently large intersite repulsion; (iii) regardless of the presence of this phase shift, superfluidity is suppressed by such {\it static} stripe formation. (iv) In contrast, at weaker coupling, the attractive fermionic Hubbard model, stripe formation is {\it not} inimical to pairing -- quantum {\it fluctuating} stripes are central to coexistence with superconductivity. In addition, the simulations' accuracy allows us to quantify the energy differences of the competing phases precisely, something which is often referred to but not commonly evaluated.

\paragraph{Model.---} Our starting point is the attractive Fermi-Hubbard model~\cite{Scalettar1989, Moreo1991, Paiva2004, Fontenele2022}
\begin{align}
    \hat {\cal H} = -t\sum_{\langle i,j\rangle,\sigma}\hat c_{i\sigma}^\dagger \hat c_{j\sigma}^{\phantom{\dagger}} + U\sum_{i}\hat n_{i\uparrow}\hat n_{i\downarrow} -\mu\sum_{i\sigma}\hat n_{i\sigma}\ ,
    \label{eq:fermion_ham}
\end{align}
where $\hat c_{i\sigma}^{\phantom{\dagger}}(\hat c_{i\sigma}^\dagger)$ annihilates (creates) an electron with spin $\sigma$ in the site $i$ of an $L_x\times L_y$ lattice and $\hat n_{i\sigma}$ is the corresponding number operator; $t$ gives the hopping integral with onsite interactions $U<0$, and the chemical potential $\mu$ regulates the fermionic density. For $|U|/t \gg1$, the pairs induced by the attractive interactions turn local, describing a composite fermion satisfying the hard-core constraint:  second-order perturbation theory recasts this Hamiltonian in terms of repulsive hard-core bosons in the presence of a rescaled chemical potential~\cite{Micnas1990}, $\hat {\cal H}^\prime = -\frac{2t^2}{|U|}\sum_{\langle i,j\rangle} \hat b_i^\dagger \hat b_j^{\phantom{\dagger}} + \frac{2t^2}{|U|} \sum_{\langle i,j\rangle} (\hat n^\prime_{i} - 1/2)(\hat n^\prime_{j} - 1/2) +(|U|-2\mu)\sum_i(1-\hat n^\prime_{i})$, where $\hat b_i^{\phantom{\dagger}}(\hat b_i^\dagger)$ is the $i$-th site annihilation (creation) operator satisfying $[\hat b_i^{\phantom{\dagger}},\hat b_j^\dagger]=0$ if $i\neq j$, while $\{\hat b_i^{\phantom{\dagger}}, \hat b_i^\dagger\}=1$, with the constraint $\hat b_i^{\dagger 2} = \hat b_i^2 = 0$~\cite{Lieb1961}; $\hat n_i^\prime = \hat b_i^\dagger\hat b_i^{\phantom{\dagger}}$ is the associated number operator. This Hamiltonian for repulsively interacting hard-core bosons, generalized to allow the hopping amplitude to differ from the nearest-neighbor interaction strength, has been extensively studied~\cite{Scalettar1995, Hebert2001, Schmid2002, Sengupta2005, Wessel2005, Zhu2020}. With only nearest-neighbor interactions in a square lattice, the possible phases are either ordered solids (with different arrangements depending on the density $\langle \hat n^\prime\rangle$) or a superfluid phase; their coexistence, i.e., a supersolid, has been ruled out~\cite{Scalettar1995, Hebert2001}. At the first-order phase transition between the quantum solids and the superfluid phase with $\langle \hat n^\prime \rangle$ close to 1/2, instability toward phase separation emerges, signaled by discontinuities in the equation of state [inset in Fig.~\ref{fig:fig_1}(f)] and characterized by domain walls separated by antiphased checkerboard patterns~\cite{Sengupta2005}. This stripe formation can be explained via the contribution to the kinetic energy gain in second-order perturbation theory in the atomic limit~\cite{Sengupta2005, Zhu2020} [see Fig.~\ref{fig:fig_1}(b)].

\begin{figure}[t]
    \centering
    \includegraphics[width=0.94\columnwidth]{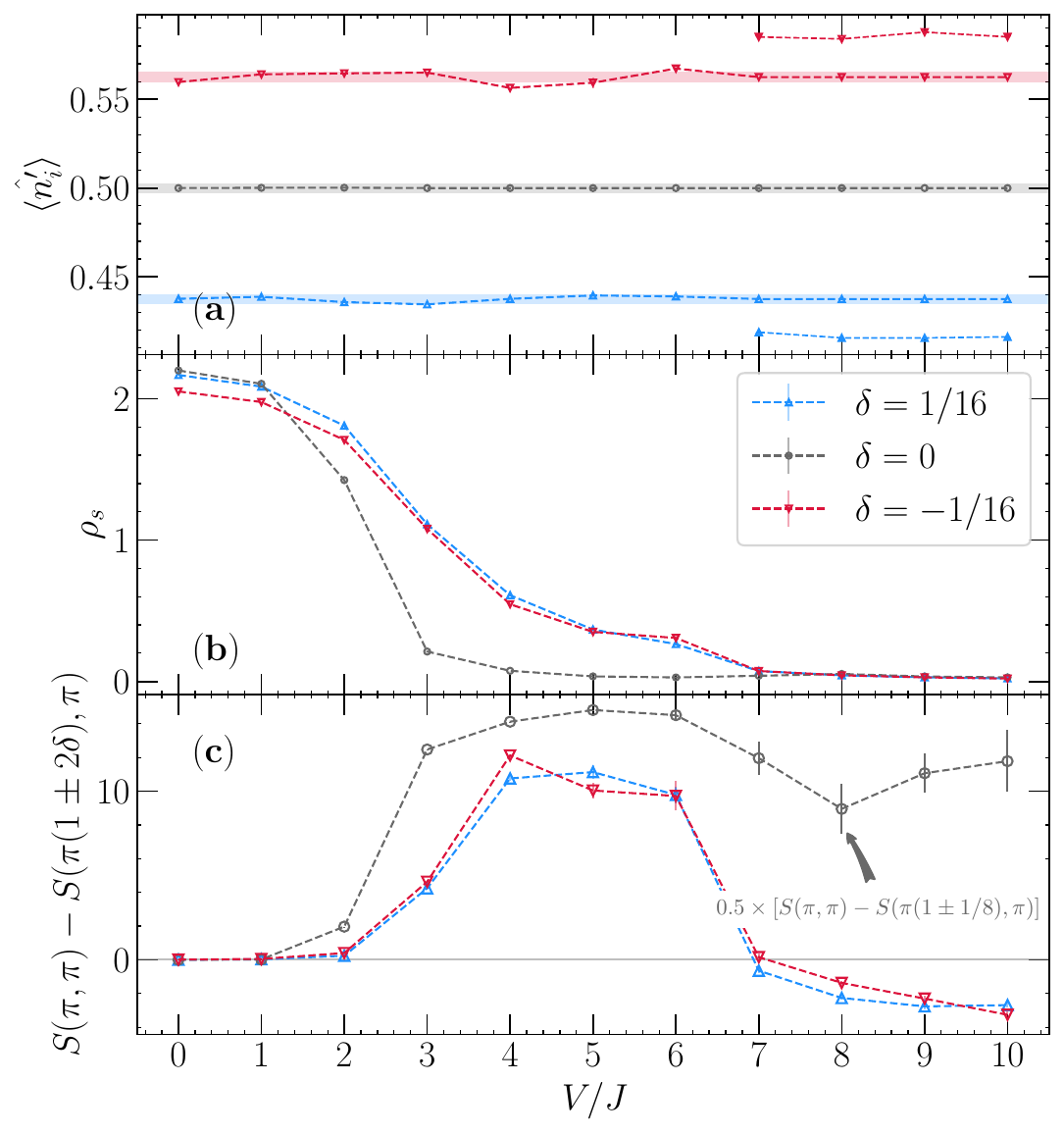}
    \caption{(a) Average density vs.~interaction strength $V$ for three doping values $\delta = 0, \pm 1/16$ on the hardcore boson model; the shaded horizontal area gives the average total density target, and the markers the calculated density -- once charge stripes are formed, these lead to a second lower (higher) density for $\delta = 1/16$ ($\delta = -1/16$) at certain regions. (b) The superfluid weight shows that in the presence of stripes, $\rho_s \to 0$. (c) The difference in the density structure factor for two values of $k_x$, $\pi$ and $\pi\pm2\delta$; when negative, it signals the emergence of stripes with anti-phase density ordering --- the undoped ($\delta = 0$) case is rescaled by 0.5 for easier visualization.}
    \label{fig:fig_2}
\end{figure}

\paragraph{Stripes in hardcore boson ladders.--} We start by characterizing the interplay between superfluidity and the emergence of stripes in ladder geometries. By breaking lattice rotational symmetry, they facilitate the manifestation of charge stripes along the short direction~\cite{Chang2010, Zheng2016, Huang2017, Boxiao2017, Huang2018, Xu2022, Huang2023, Xu2023, Jiang2024}. The Hamiltonian reads 
\begin{align}
    \hat {\cal H}_{\rm hcb}=-J\sum_{\langle {i,j} \rangle } (\hat b_i^\dagger \hat b_j^{\phantom{\dagger}} + \hat b_j^\dagger \hat b_i^{\phantom{\dagger}} )
+V \sum_{\langle i,j \rangle } \hat n_i^\prime \hat n_j^\prime -\mu \sum_i\hat n_i^\prime\ ,
\end{align}
where we use the stochastic series expansion quantum Monte Carlo method~\cite{Sandvik1999, Alps02, Alps03} at sufficiently low temperatures $T$ by setting $\beta \equiv 1/T = 2L_x$. Convergence is assisted towards the formation of stripes when including a pinning staggered potential at the edges of the system~\cite{Boxiao2017, Xu2022, Xu2023}, $h(-1)^{(i_x+i_y)}\hat n^\prime_i$ for $i_x = 0$ and $L_x-1$ --- see Fig.~\ref{fig:fig_1}(a). Finally, we adjust the chemical potential $\mu$ such that the hole-doping $\delta \equiv 1/2 - \langle \hat n^\prime\rangle \simeq 1/16$. In the original picture of spinful fermions, this would correspond to the doping $\delta=1/8$, often studied in the context of stripe formation in the repulsive Hubbard model~\cite{Boxiao2017, Huang2023} and where the `stripe anomaly' emerges in certain classes of cuprates (i.e., where stripes are most robust while suppressing bulk superconductivity~\cite{Hucker2011, Tranquada2020}).

\begin{figure}[t]
    \centering
    \includegraphics[width=1.0\columnwidth]{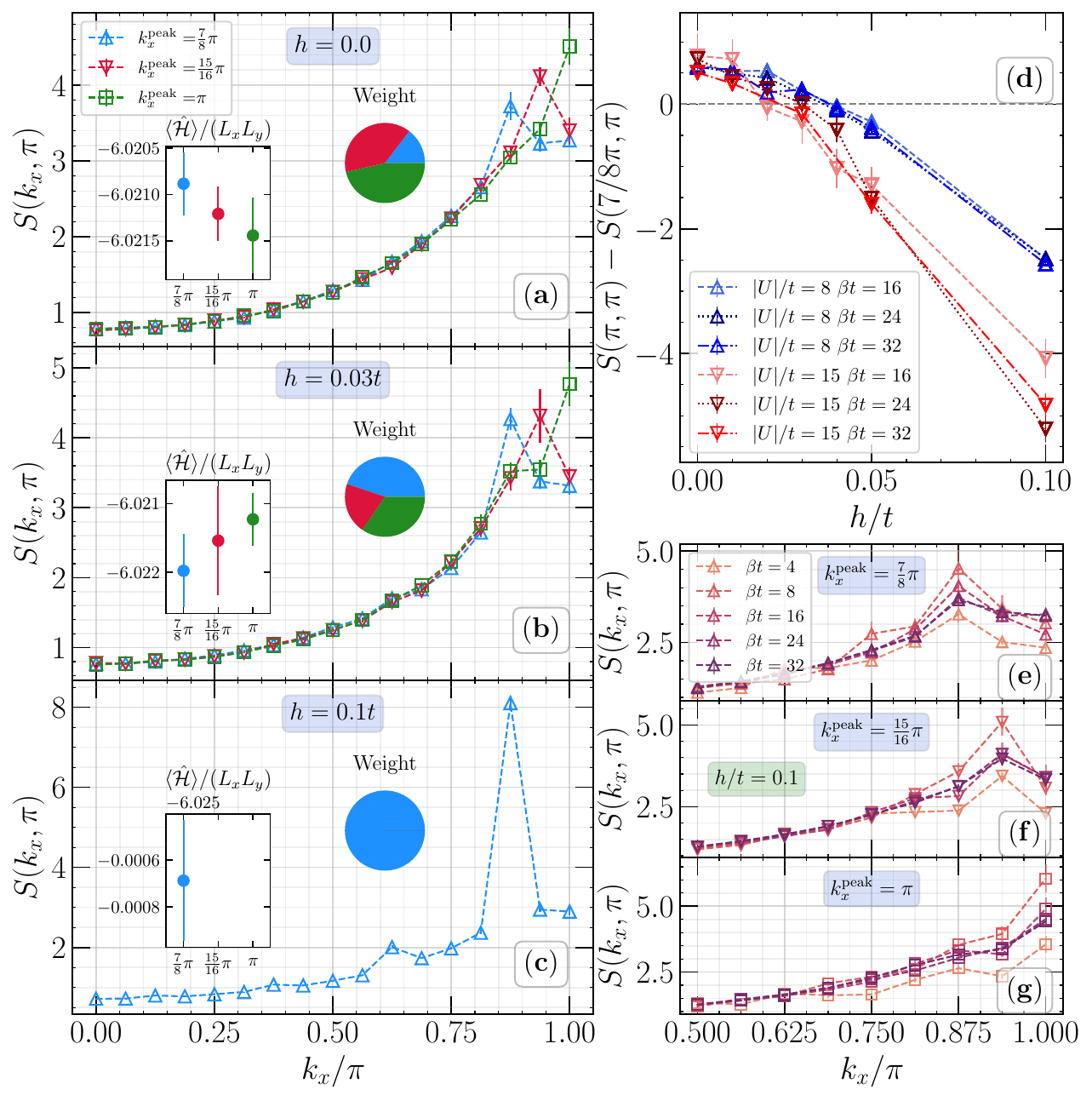}  
    \caption{(a--c) Structure factor $S(k_x,\pi)$ for the \textit{fermionic} model at $\beta t = 24$ with increasing magnitude of the modulated pinning potential $h/t = 0; 0.03$ and 0.1, respectively; here the realizations are filtered according to the peak location $k^{\rm peak}_x$ --- the pie chart gives the fraction of the total number of realizations exhibiting a given peak in $S(k_x,\pi)$ and the remaining inset the average energy per site for each corresponding $k_x$-value.
     (d) The difference between the CDW structure factor and the one that gives a period-8 stripe averaged over \textit{all} realizations: only after imposing a large pinning potential can a robust stripe be formed. 
      (e--g) The temperature dependence of $S(k_x,\pi)$ with $h/t=0.1$ for realizations exhibiting $k_{\rm peak} = \frac78\pi, \frac{15}{16}\pi$ and $\pi$, respectively. Data are extracted on a $32\times 4$ torus with $|U|/t=15$ when not indicated otherwise.
     %  \tcr{What is the point of panels (e-g)?  I am seeing the peaks at first 
     % increasing as $\beta$ becomes larger, but then they appear to come back down? We do not explicitly refer to these panels in the main text, though we do say that the interaction strengths
     % are not sufficient to induce a checkerboard solid.  We say $h=0.1$ for these panels in the caption.  Should we also put it in their legends for emphasis?}
     } 
    \label{fig:fig_3}
\end{figure}

The site-resolved average density is shown in Fig.~\ref{fig:fig_1}(c) on a $32\times 4$ lattice with open (periodic) boundary conditions in the long (short) direction at the strongly interacting regime with $V/J=8$, making immediately apparent the regions with increased hole density, periodically modulated along the ladder. Additionally, charge stripes separate regions where robust checkerboard patterns emerge. These two aspects can also be readily seen in the average staggered hole density, $(-1)^{i_x+i_y} (1/2-\langle \hat n^\prime\rangle)$, along the ladder [Fig.~\ref{fig:fig_1}(d)] showing the region of anti-phase, which is separated by hole-rich stripes wherein a node of the staggered hole density emerges. The decay profile of the two-point correlations,  $c_{\bf i_0}(i_x) = \langle \hat n^\prime_{{\bf i}}\hat n^\prime_{\bf i_0}\rangle - \langle \hat n^\prime_{{\bf i}}\rangle\langle \hat n^\prime_{\bf i_0}\rangle$ (along $i_y=1$), complements this analysis where the reversal of the staggered periodicity is observed at the stripe locations -- see vertical lines in Fig.~\ref{fig:fig_1}(e).

To quantify precisely the characteristic stripe wavelength, we compute the structure factor of the two-point density correlations,
\begin{equation}
    S(\mathbf{k}) = \frac{1}{L_xL_y}\sum_{i,j,\mathbf{k}}e^{{\rm i}\mathbf{k}\cdot\left( \mathbf{r}_i - \mathbf{r}_j\right)}\left\langle \left( \hat n_i^\prime - \langle \hat n_i^\prime \rangle \right) \left( \hat n_j^\prime - \langle \hat n_j^\prime \rangle\right)\right\rangle\ ,
    \label{eq:struct_fac}
\end{equation}
where we focus on the momenta $\mathbf{k} = (k_x,\pi)$. In the undoped case ($\delta = 0$), this quantity peaks at $\mathbf{k} = (\pi,\pi)$ owing to the robust checkerboard solid that emerges at such strong values of the nearest-neighbor interactions.  In the presence of finite doping, we observe that its peak is now displaced to $k_x = \pi (1\pm2\delta)$ [Fig.~\ref{fig:fig_1}(f)], a direct signature of the stripe formation and the antiphase the checkerboard domains exhibit across stripes~\cite{Wietek2021, Schlomer2022}. Notably, for this case of hole-doping with $\delta = 1/16$, even without edge pinning potentials ($h=0$), the $\pi$-phase striped charge density wave (CDW) is robust. Hardcore-boson doping also leads to similar behavior while typically requiring larger pinning potentials -- see Supplementary Materials (SM)~\cite{SM}. Conversely, the stripe behavior is reversed: \textit{density-rich} regions separate anti-phase $\langle \hat n^\prime\rangle \simeq 1/2$ checkerboard solid domains instead -- see SM~\cite{SM} for a system-size analysis.

Figure~\ref{fig:fig_2} shows how $V/J$ affects this picture. The density inhomogeneity associated with the stripes is only seen for large values of $V/J$ [Fig.~\ref{fig:fig_2}(a)], a signature that just when doping the $\langle \hat n^\prime\rangle=1/2$ checkerboard solid, originally formed at $V/J\geq2$ in the  $\delta = 0$ regime, one may then observe charge stripes --- this is the case for both hole ($\delta >0$) and hardcore boson doping ($\delta <0$). Figure~\ref{fig:fig_2}(c) shows that a robust density $\pi$-phase shift, i.e., $S(\pi,\pi)-S(\pi(1\pm2\delta),\pi)<0$, is intimately tied to the stripe formation at large interactions. Finally, we notice that increasing $V/J$, inducing the checkerboard solid formation, suppresses the superfluidity [Fig.~\ref{fig:fig_2}(b)], quantified by $\rho_s = \langle W_x^2 + W_y^2\rangle/(2\beta J)$, where $W_{x(y)}$ is the winding number of the bosonic world-lines in $x$($y$) directions~\cite{Pollock1987}. The formation of sublattice reversal of checkerboard domains does not change this picture- a finite superfluid density is incompatible in this case as well. If enforcing the stripe emergence by an externally imposed  potential~\cite{Mondaini2012, Ying2022, Chen2024} such anti-correlation between finite superfluidity and a robust manifestation of a $\pi$-phase shift persists -- see SM~\cite{SM}. 

\paragraph{Stripes in $U<0$ Fermi-Hubbard ladders.--} Having established that stripes naturally emerge in a model for hardcore bosons and that these compete with the superfluid properties once CDW domains are formed (with antiphase or not), we tackle the spinful attractive Hubbard model~\eqref{eq:fermion_ham} to investigate if similar conclusions carry over. While its strongly interacting regime with increasingly local pairs is suggestive that similar results of the hardcore boson case would emerge, we argue below that much stronger quantum fluctuations lead to very different outcomes. In particular, the mapping to the hardcore boson Hamiltonian shows that in approaching the $|U|/t \gg1$ limit, the corresponding hardcore boson model displays interaction strengths ($V=J$) which are not sufficient to induce a checkerboard solid at $\delta=0$.  
% \tcr{See red comments in Figure 3 caption.  Is the preceding statement based on the non-monotonic $\beta$ dependence in panel (g)?  It seems true also of panels (e,f).}

\begin{figure}[t]
    \centering
    \includegraphics[width=1.0\columnwidth]{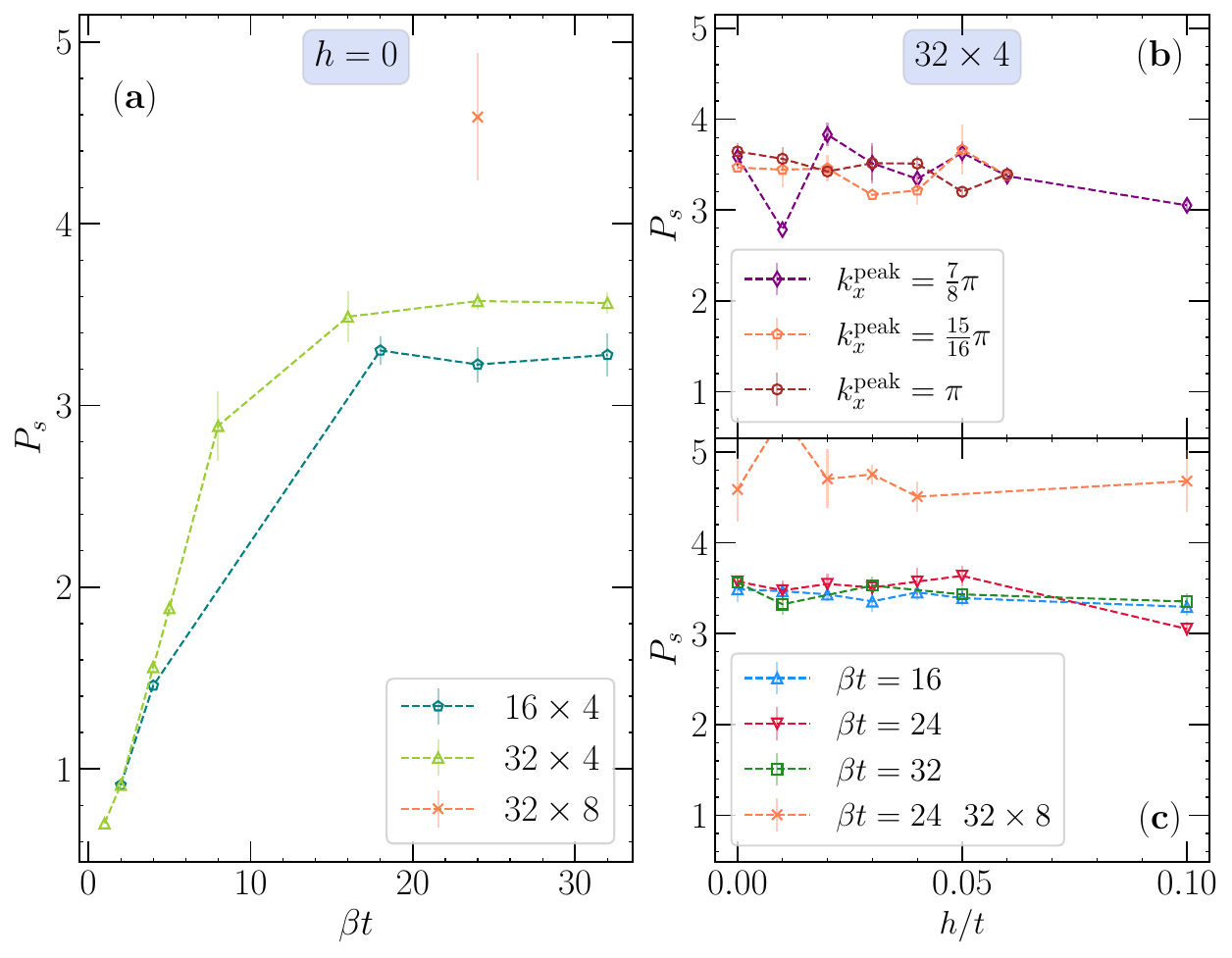}  
    \caption{(a) Saturation of the $s$-wave pair structure factor with the inverse temperature for two different system sizes without antiphase staggered pinning potentials $h(-1)^{i_x+i_y + i_x/8}\hat n_i$. The effect of $h$ on the pairing for the $32\times 4$ lattice, resolved by $k_x^{\rm peak}$ (b) or by temperature and system size (c). Data are shown for $|U|/t=15$; imaginary-time discretization is $t\Delta\tau = 0.05$ as in Fig.~\ref{fig:fig_3}.} 
    \label{fig:fig_4}
\end{figure}

We start by showing in Fig.~\ref{fig:fig_3} the density structure factor [Eq.~\eqref{eq:struct_fac} with density operators for the fermionic case] with hole-doping $\delta = 1 - \langle \hat n\rangle \simeq 1/8$ and $|U|/t = 15$ on the same $32\times 4$ ladder at $\beta t = 32$. We focus on a torus geometry (periodic boundaries in both directions) with an \textit{antiphase} staggered potential $h(-1)^{i_x+i_y + i_x/8}\hat n_i$ applied at $i_x = 0, 8, 16, 24$. Imposition of a pinning potential provides a {\it useful tool} to filter antiphase configurations and assess their effect on pairing. Here, in particular, the competition of states that emerge at these low temperatures becomes clear: The simulations statistically converge such that $S(k_x,\pi)$ exhibits peaks at $k^{\rm peak}_x = \pi$, $15\pi/16$, or $7\pi/8$ and their relative weight [i.e., the fraction of independent Markov chains (realizations) that converge to show a given peak in $S(k_x,\pi)$] can be tuned according to the magnitude of the pinning potential $h$, see Fig.~\ref{fig:fig_3}(a--c). The corresponding average energy per site $\langle \hat {\cal H}\rangle/(L_xL_y)$ (insets) explicitly depicts this competition, demonstrating that they are energetically very close.  

Using $\Delta S \equiv S(\pi,\pi) - S(7\pi/8,\pi)$ as a proxy of robust antiphase period-8 stripe formation, Fig.~\ref{fig:fig_3}(d) shows the necessity of including a pinning potential to tip the balance in favor of a stripe state with $\Delta S < 0$, also for smaller values of the attractive interactions strengths $|U|$ and temperatures. Focusing on the latter, we notice that in resolving $S(k_x,\pi)$ by the corresponding peak location across different realizations [Fig.~\ref{fig:fig_3}(e--g)], one can see relatively resilient peaks away from $k_x^{\rm peak}\neq \pi$ even at $T/t = 4$ for $h/t = 0.1$.  This bodes well for observation in current experiments of trapped cold atoms emulating the attractive Hubbard model that has been shown to tackle a similar temperature range~\cite{Hartke2022}.

Lastly, we focus on the interplay of stripes and pairing, here quantified by the zero-momentum $s$-wave pair structure factor $P_s = \sum_{i, j}\langle \hat \Delta_i^{\phantom{\dagger}} \hat \Delta_j^\dagger\rangle/(L_xL_y)$, where $\hat \Delta_i = \hat c_{i\uparrow}\hat c_{i\downarrow}$ annihilates a fermionic pair at site $i$. Low temperatures lead to a saturation of this quantity that is extensive with the system size [Fig.~\ref{fig:fig_4}(a)], indicating an expected pairing long-range order when approaching the ground-state~\cite{Scalettar1989}. While these results in the absence of pinning potential are not a guarantee that stripes are necessarily influencing them [see Fig.~\ref{fig:fig_3}(a)], only a negligible decrease of the pairing is observed once stripe modulations are induced via the anti-phase pinning potential at low-temperatures [Fig.~\ref{fig:fig_4}(c)], even if explicitly resolving the realizations that exhibit different $k_x^{\rm peak}$ in $S(k_x,\pi)$ [Fig.~\ref{fig:fig_4}(b)]. This poses a contrast to the case of hard-core bosons, where stripes and superfluidity were incompatible, hinting that the larger quantum fluctuations when the Cooper pairs are not made local allow one to see the concomitant manifestation of pairing and phase flip density modulations. 

\paragraph{Summary and outlook.--} Disentangling the interplay of density modulations and pairing is at the core of understanding the physics of high-temperature superconductors. Here, we showed that stripes with antiphase checkerboard density are inimical to the superfluidity in interacting hard-core boson models: These can naturally form at strong interactions, leading to a stripe crystal that suppresses Bose-Einstein condensation. When these bosonic Cooper pairs are not tightly bound, stripe states emerge with more than one characteristic wave vector at low-$T$'s [$k_x = \pi(1\pm \delta)$ or $\pi(1\pm\delta/2)$], consistent with fluctuating-stripe proposals, and are not seen to hurt the isotropic pairing significantly. Such a picture, here shown not to be necessarily tied to the repulsive Hubbard model, enlarges the scope in which multiple orders intertwine, opening the prospects of its observation in trapped ultracold atom experiments with controlled attractive interactions~\cite{Mitra2018, Brown2020, Chan2020, Hartke2022} where correlations of the {\it attractive} Hubbard model at different interaction strengths already identify regimes of tightly bound pairs. The accessible temperature range
is close to the values that we show here are sufficient to see a $\pi$-phase shift across stripes. Systematic analysis of the correlations could provide signals of $\pi$ ordering, as well as peaks at $\pi \pm 2\delta$ we find in Fig.~\ref{fig:fig_1}(f).

\begin{acknowledgments}
\paragraph{Acknowledgments.---} T.Y.~acknowledges support of the Natural Science Foundation of Heilongjiang Province (Grant No.~YQ2023A004). R.M.~acknowledges support from the NSFC Grants No.~NSAF-U2230402 and No.~12222401. R.T.S.~is supported by the grant DOE DE-SC0014671 funded by the U.S. Department of Energy, Office of Science. Numerical simulations were performed in the Tianhe-2JK at the Beijing Computational Science Research Center.
\end{acknowledgments}

\bibliography{references}

%%%%%%%%%%%%%%%%%%%%%%%%%%%%%%%%%%%%%%%%%%%%%%%%%%%%%%%%%%%%%%%%
%%%%%%%%%%%%%%%%%%%%%%%%%%%%%%%%%%%%%%%%%%%%%%%%%%%%%%%%%%%%%%%%
\beginsupplement
%%%%%%%%%%%%%%%%%%%%%%%%%%%%%%%%%%%%%%%%%%%%%%%%%%%%%%%%%%%%%%%%
%%%%%%%%%%%%%%%%%%%%%%%%%%%%%%%%%%%%%%%%%%%%%%%%%%%%%%%%%%%%%%%%

\clearpage

\renewcommand{\theequation}{S\arabic{equation}}
\renewcommand\thesection{\arabic{section}}
\renewcommand{\thetable}{S\arabic{table}}%
\renewcommand{\thefigure}{S\arabic{figure}}%

\setcounter{equation}{0}

\onecolumngrid

\begin{center}

{\large \bf Supplementary Materials:
 \\ Stripes and the Emergence of Charge $\pi$-phase Shifts in Isotropically Paired Systems}\\

\vspace{0.3cm}

\end{center}

\vspace{0.6cm}

\twocolumngrid

In these Supplementary Materials, we provide additional data to support the findings in the main text, including an analysis of superfluidity with externally imposed stripes in the hard-core boson model, results for different system sizes, and stripe formation with an excess of hard-core bosons. Lastly, we show the momentum dependence of the pairing correlations for the fermionic case.

\section{`Negative' doping in the hardcore bosons case} 
\label{sec:neg_doping}
\begin{figure}[!h]
    \centering
    \includegraphics[width=1.0\columnwidth]{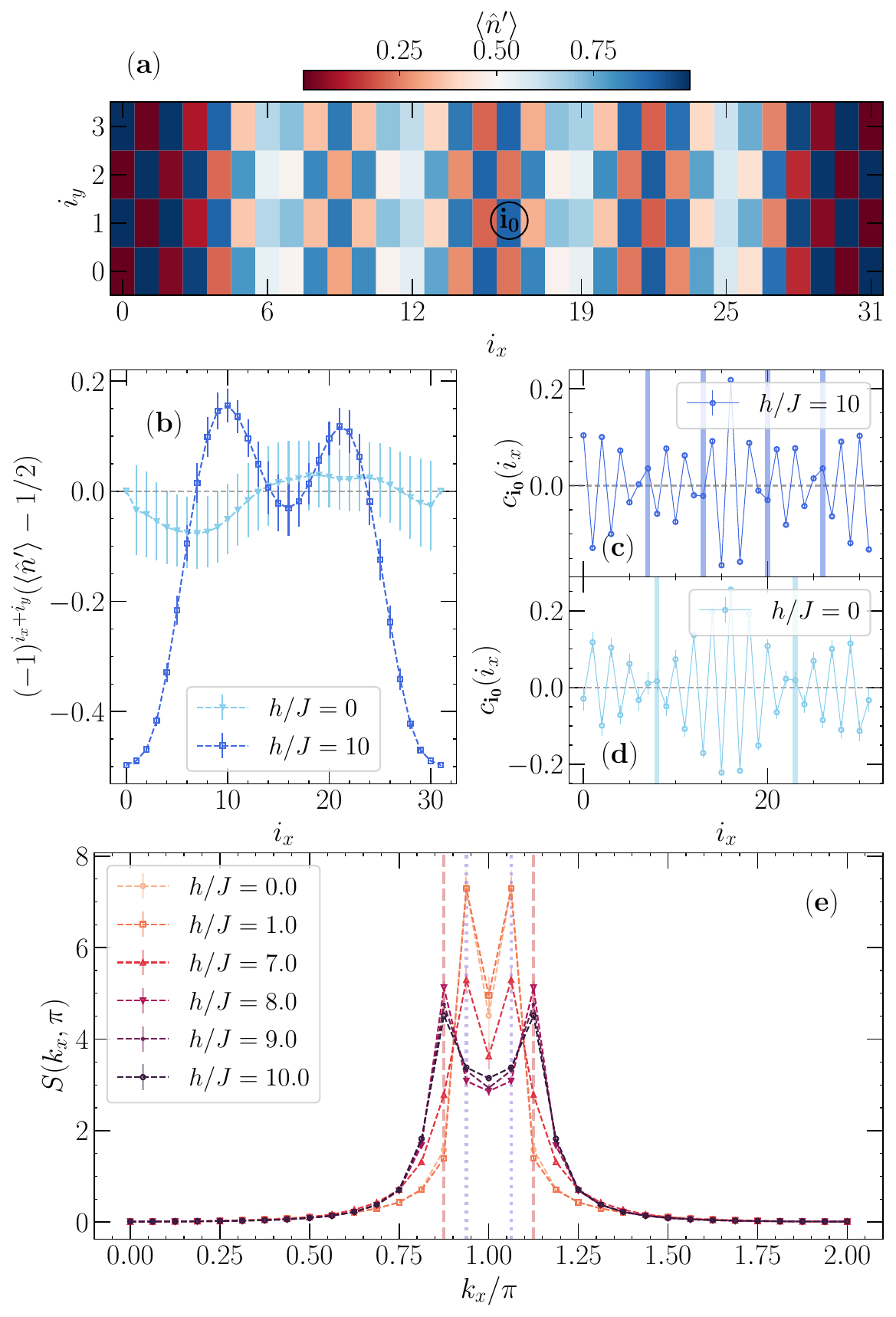}
    \caption{(a) Site-resolved density in a $32\times 4$ cylinder doped to render an averaged density $\langle \hat n^\prime\rangle = 0.5625$ for the hardcore-boson Hamiltonian with $V/J = 8$ and edge pinning staggered potential $h/J=10$. (b) The profile of the staggered excess density with ($h/J=10$) and without ($h/J=0$) a pinning potential and (c) and (d) the corresponding decay profile of the two-point correlation functions [measured in relation to $\bf i_0$ marked in (a)]. (e) Structure factor of density-correlations $S(k_x,\pi)$ vs.~$k_x$ for various values pinning potential amplitude: for $h/J \lesssim 7$ the peaks are at $k_x^{\rm peak} = \pi(1\pm\delta)$, a signature of two bulk stripes, whereas for larger pinning potentials the peak $k_x^{\rm peak} = \pi(1\pm2\delta)$ is characteristic of four stripes.}
    \label{fig:SM_Fig1}
\end{figure}

In the main text, we focus on the formation of stripes, described by the periodic modulation of the extra holes and the sublattice reversal of the charge order across one such hole-rich region. Here, we will show that similar phenomenology can be obtained when one dopes the half-filling $\langle \hat n^\prime\rangle=1/2$ with extra particles. In particular, we focus on the density of $\langle \hat n^\prime\rangle = 1/2 + 1/16 = 0.5625$ where another discontinuity at the $\langle \hat n^\prime\rangle$ vs.~$\mu$ curve emerges [inset in Fig.~\tcr{1}(f)], a characteristic of phase separation~\cite{Sengupta2005}, of which the stripes are an example. Figure~\ref{fig:SM_Fig1} gives a variant of Fig.~\tcr{1} of the main text for this different doping.

The density distribution with a large staggered edge pinning field $h/J=10$, Fig.~\ref{fig:SM_Fig1}(a), exhibits stripe formation but with quasi-one-dimensional regions displaying instead \textit{excess} charges separating regions where antiphase checkerboard pattern materializes. Their location is given by the nodes of the staggered excess density, $(-1)^{i_x+i_y}(\langle \hat n^\prime_i\rangle - 1/2)$ [Fig.~\ref{fig:SM_Fig1}(b)], or by the region in which a reversal of the two-point density correlations $c_{\bf i_0}(i_x)$ change its staggered pattern along the ladder [Fig.~\ref{fig:SM_Fig1}(c)] -- here we take correlations along $i_y = 1$. Notably, without a pinning potential, the staggered excess density is no longer a proxy for stripe formation, but this is obviously not an issue for the two-point correlations in Fig.~\ref{fig:SM_Fig1}(d). Using the reversal of the staggered correlation pattern as the signature of phase flip lines, we infer the presence of just two stripes in this case, in contrast to four for $h/J\lesssim 7$ in the bulk of the ladder. This results in different peak locations of the density structure factor $S(k_x,\pi)$, at $k_x^{\rm peak} = \pi(1\pm2\delta)$ in the case of four stripes, and $k_x^{\rm peak} = \pi(1\pm\delta)$ for two stripes, see Fig.~\ref{fig:SM_Fig1}(e). The latter, however, as apparent by the staggered excess density, exhibits pinned values at zero for $i_x = 0$ and $L_x-1$, indicating that the open edges are `sinks' for the excess particles doped in the system for small or vanishing pinning potentials.

An interpretation of these results is that in the strongly interacting regime ($V/J=8$), energy minimization is achieved by forming an extended $\langle \hat n^\prime\rangle = 1/2$ solid; the added particles thus tend to accumulate at the open edges of the cylinder (that is, with periodic boundary conditions along the short direction with open ones on the long direction). To observe robust stripe formation, the edge pinning potentials must be substantially large to expel them from the lattice's ends so they will condense periodically along the ladder, leading to a two-to-four stripe pattern transition. This is to be contrasted with the hole doping in the main text for the same lattice size, where four period-8 stripes are seen irrespective of the $h/J$ value.

\section{Finite-size effects for the hole-doped hardcore boson model}

\begin{figure}[!t]
    \centering
    \includegraphics[width=1.0\columnwidth]{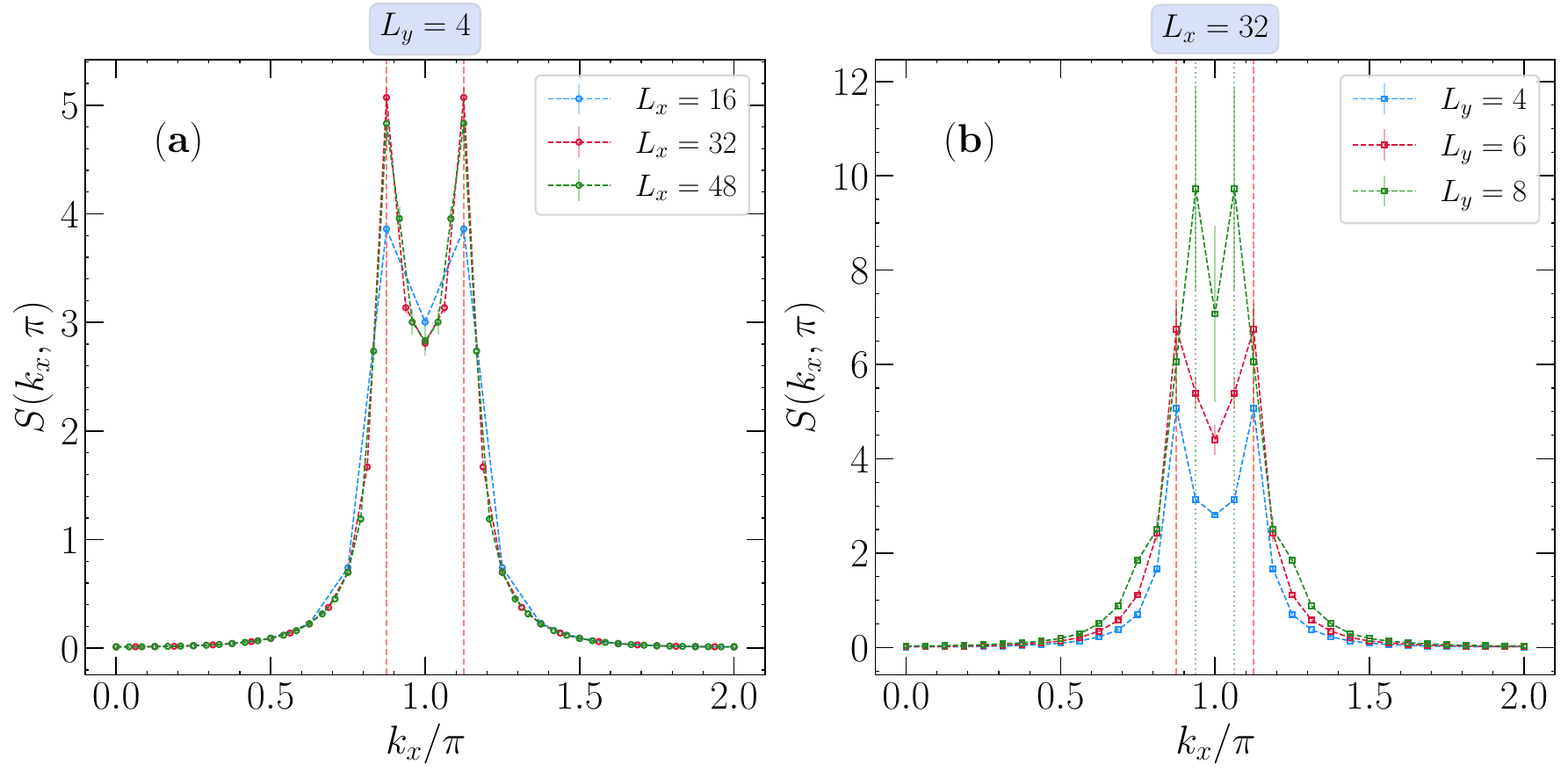}
    \caption{Density structure factor $S(k_x,\pi)$ vs.~$k_x$ for the hardcore boson model with fixed $L_y = 4$ and various lengths $L_x$ in (a) and growing width $L_y$ for a fixed $L_x=32$, (b). Here, the parameters are $V/J=8$, the staggered pinning potential amplitude is $h=J$, and the hole doping is $\delta = 1/16$. Red dashed [green dotted] vertical lines describe the peak position at $k_x^{\rm peak} = \pi(1\pm2\delta)$ [$k_x^{\rm peak} = \pi(1\pm\delta)$].}
    \label{fig:SM_Fig2}
\end{figure}

Our results have so far demonstrated that the formation of either hole-rich (in the main text) or density-rich (here in the previous section) regions can exhibit the phenomenon of sublattice reversal of the charge ordering in relatively small lattices featuring $32\times 4$ sites. Our goal now is to extend this analysis performed in the strongly interacting regime ($V/J=8$) to larger lattices by monitoring the density-density structure factor $S(k_x,\pi)$, as originally seen in the main text. First, we notice that in applying a staggered density pinning field with magnitude $h=J$, for fixed-width ladders ($L_y=4$) with various lengths $L_x$ [Fig.~\ref{fig:SM_Fig2}(a)] at densities $\langle \hat n^\prime\rangle = 0.4375$, the peak at $k_x = \pi(1 \pm 2\delta)$ is robustly consistent in lattices ranging from $L_x = 16$ to 48, signifying period-8 stripe formation with anti-phase of the checkerboard density patterns. 

Next, we analyze the effect of the ladder's width at a fixed $L_x = 32$ at the same density. While the original case with $L_y = 4$ shows robust peaks in $S(k_x = \pi(1\pm 2\delta),\pi)$, this is not as clear for wider ladders. For $L_y = 6$, these `shoulder-peaks' start competing in magnitude with the one at $k_x = \pi$ (full checkerboard solid), where again, the large error bars in the latter indicate competing orders. A similar situation occurs in even wider ladders with $L_x = 8$: here, the main peaks are in $k_x = \pi(1\pm \delta)$ instead. In this largest case, some solutions converge to the absence of stripes, i.e., with a main $k_x = \pi$ peak; note, however, the large error bars that stem from the fact that only a limited subset of realizations acquired this $(\pi,\pi)$-solid charge ordering. Overall, this analysis exemplifies that the actual ground state (and the ensuing stripe periodicity) exhibits strong dependency on the ladder geometries, and even if tipping the balance with a pinning staggered potential, it might not be sufficient to select one type of stripe ordering with anti-phase in the checkerboard density pattern. Such behavior, strong sensitivity of the ground states to system sizes, has also been seen in the context of the repulsive Hubbard ladders using density matrix renormalization and constrained path quantum Monte Carlo~\cite{Xu2023}, an indication of the presence of competing states that are very close in energy.

\begin{figure}[!t] 
  \includegraphics[width=0.6\columnwidth]{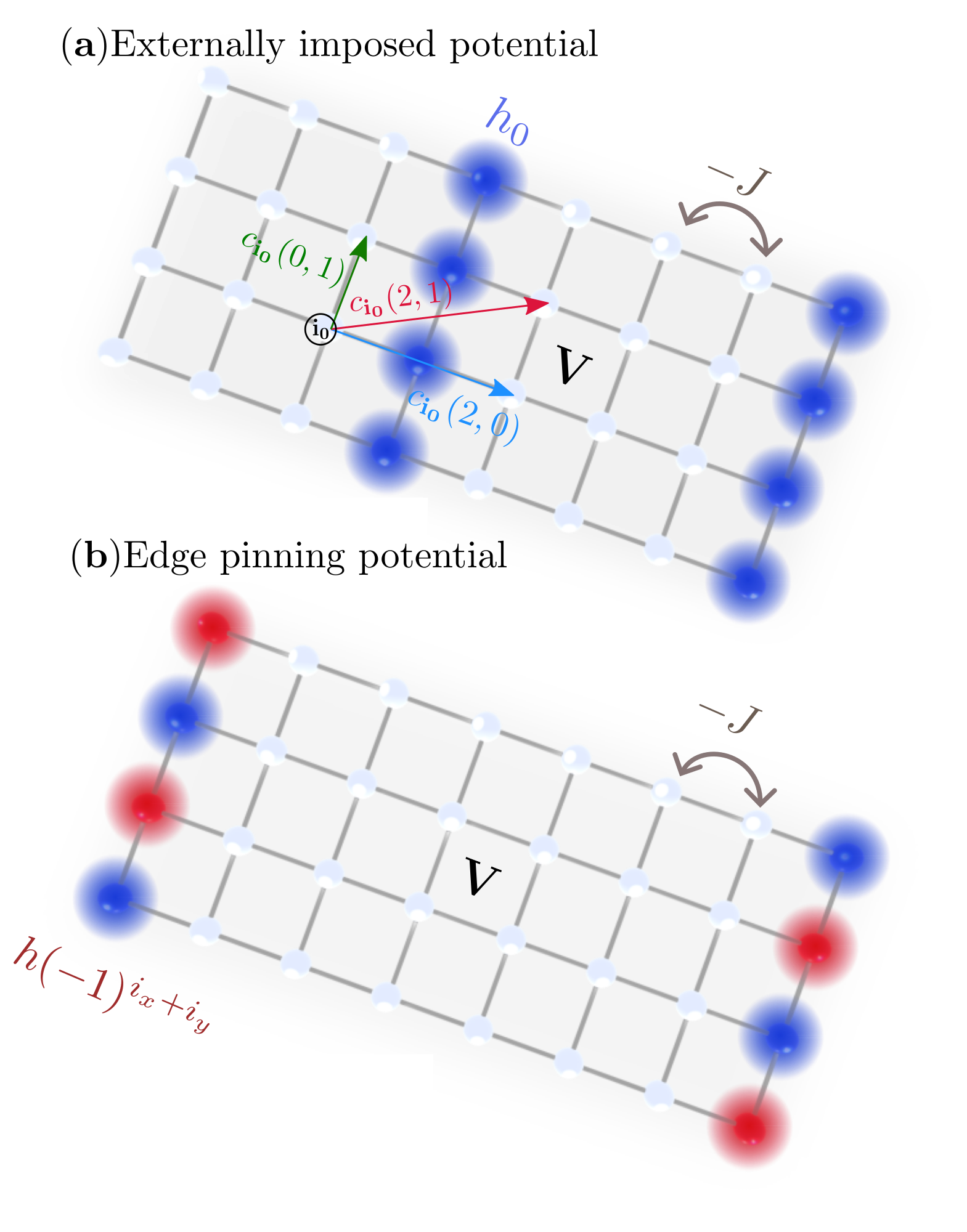}
   \caption{Cartoons contrasting the two types of hard-core boson Hamiltonians we study, with (a) a periodically imposed external potential and (b) with an edge pinning potential, initially used in the main text. Here, the $8\times 4$ ladder is the one used in the ED calculations. In (a), we also annotate the two-point correlation distances used in the remaining of this section.}
   \label{fig:SM_cartoon}
\end{figure}

\begin{figure*}[t]
   \includegraphics[width=1.9\columnwidth]{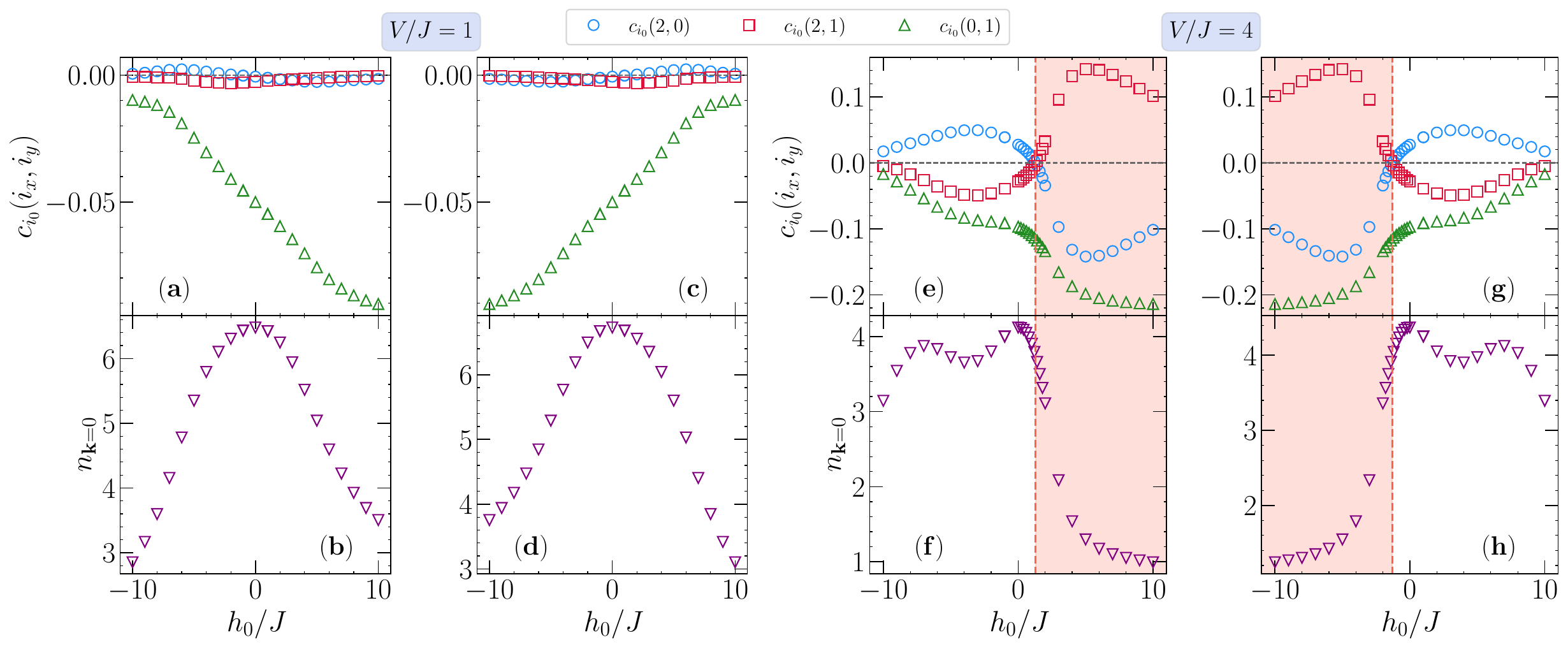}
   \caption{ED results for an $8\times 4$ lattice with stripe periodicity ${\cal P} = 4$, and interactions $V/t = 1$ (a--d). The left (a, b) and right (c, d) panels show results for $\langle \hat n^\prime\rangle = 3/8$ and $5/8$, respectively. Both the correlations $c_{i_0}(i_x,i_y)$ and the zero-momentum occupancy $n_{{\bf k} = 0}$ are shown as a function of the stripe chemical potential $h_0$. Note the symmetry $h_0 \longleftrightarrow -h_0$ with $\langle \hat n^\prime\rangle \longleftrightarrow 1 - \langle \hat n^\prime\rangle$ in the data. (e--h) The same as (a--d), but for repulsive interactions $V/t = 4$. Here, the $\pi$-phase shift is easily observed (shaded regions) and is accompanied by a systematic reduction in the zero-momentum occupancy.}
   \label{fig:ED_result_V1_4}
\end{figure*}

\section{Externally imposed stripes}
Up until now, we have investigated the spontaneous formation of charge stripes in the case of repulsively interacting hard-core bosons or locally attractive fermions. Such a `stripe crystal' manifestation is facilitated by imposing a small pinning field, which triggers the density-wave pattern. Here, we will tackle a different approach: The stripes are externally imposed via an onsite potential. That is, one-dimensional regions with reduced/enhanced density can be enforced by regulating a phenomenological chemical potential $h_0$ over periodically chosen sites in the lattice, an approach that has been previously used in studies of the repulsive~\cite{Mondaini2012, Chen2024, Chen2024b} or attractive~\cite{Ying2022} Hubbard model to infer the interplay of stripes and pairing amplitudes.
In this case, the hard-core boson Hamiltonian reads
\begin{equation}
  \hat {\cal H}=-J\sum_{\langle {i,j} \rangle } (\hat   b_{i}^\dagger \hat b_{j}^{\phantom{\dagger}} + \hat     b_{j}^\dagger \hat b_{i}^{\phantom{\dagger}} )
    +V \sum_{\langle {i,j} \rangle } \hat n_i^\prime \hat   n_j^\prime
    + h_0\sum_{i_x \in {\cal P}} \hat n_i^\prime,
    \label{eq:ham}  
\end{equation}
where $\hat b_{i}^\dagger$ ($\hat b_{i}^{\phantom{\dagger}}$) creates (annihilates) a hard-core boson at site $i$ of an $L_x\times L_y$ lattice, and $\hat n_i^\prime = \hat b_{i}^\dagger \hat b_{i}^{\phantom{\dagger}}$ is the number density operator. In what follows, we take both the cases where $L_x = 8, L_y = 4$ in exact diagonalization (ED) calculations, and the  $L_x = 8, L_y = 8$ with stochastic series expansion (SSE) quantum Monte Carlo method, choosing `sufficiently' low temperatures by setting $\beta = 1/T = 2L_{x}$. The linear dimension $L_x$ is chosen as a multiple of ${\cal P}$ to accommodate a stripe with periodic boundary conditions in both directions -- we fix ${\cal P}=4$, see Fig.~\ref{fig:SM_cartoon}(a).

As has been previously demonstrated~\cite{Ying2022}, for the case of the attractive Hubbard model, a key factor for the emergence of a charge $\pi$-phase shift is that the interstripe density is close to half-filling, i.e., $\frac{1}{L_y(L_x-{\cal P})}\sum_{i\notin {\cal P}}  \langle \hat n_i^\prime\rangle \equiv \langle \hat n^\prime\rangle_{\rm dom} = 1/2$. Thus, a local charge density wave pattern can occur, which may or may not be in phase with the next one separated by a stripe region. As a result, when dealing with the grand-canonical SSE simulations in the square lattice, we focus on the case that the added global chemical potential $-\mu\sum_i \hat n_i^\prime$ is adjusted for each value of $h_0$ to obey this constraint. In turn, for the canonical ED calculations, we focus on the fillings where, in the $|h_0|\gg J$ limit, the interstripe regions would also render $\langle \hat n^\prime\rangle_{\rm dom} \to 1/2$, via the emergence of `empty' ($h_0 > 0$) or `filled' ($h_0<0$) stripes. For a $ {\cal P} = 4$ modulation in the $8\times 4$ ladder, this results in $\rho = 3/8$ and $5/8$ (12 and 20 hardcore bosons, respectively).

In both approaches, we compute, as in the main text, the density-density correlations,
\begin{equation}
    c_{i_0}(i_x,i_y) = \left\langle (\hat n_i - \langle \hat n_i\rangle)(\hat n_{i_0} - \langle \hat n_{i_0}\rangle)\right\rangle 
\end{equation}
in respect to a given site $i_0$. This site, neighboring a stripe, is annotated in Fig.~\ref{fig:SM_cartoon}(a). As an indication of the bosonic condensation, we further compute the zero-momentum occupancy, 
\begin{equation}
    n_{{\bf k} = 0} = \frac{1}{L_xL_y}\sum_{i,j} \langle \hat b^\dagger_i \hat b_j^{\phantom{\dagger}}\rangle.
\end{equation}
As pointed out in the main text, within second-order perturbation theory, the attractive Hubbard model with $|U|/t \gg 1$ is mapped onto repulsive hardcore bosons, wherein the creation and annihilation operators of hardcore bosons are related to the annihilation and creation of pairs, respectively, $\hat b_i = \hat \Delta_i^\dagger$ and $\hat b_i^\dagger = \hat \Delta_i$~\cite{Micnas1990, Robaszkiewicz1981, Mondaini2018, Jin2022}. As a result, the zero-momentum occupancy in the model of Eq.~\eqref{eq:ham} essentially maps onto the $s$-wave pair structure factor in its fermionic counterpart.

Finally, as also explained in the main text, we quantify the superfluidity in the SSE calculations through the computation of the superfluid density
\begin{equation}
    \tilde \rho_s = \frac{\langle W_x^2 + W_y^2\rangle}{2\beta J}\ ,
\end{equation}
where $W_{x,y}$ is the winding number of the bosonic world lines in $x,y$ directions.

\subsection{ED results}
We start by describing the ED results in the $8\times 4$ lattice, as shown in Figs.~\ref{fig:ED_result_V1_4} for both weak ($V/J=1$) and strong ($V/J=4$) repulsive interactions.
First one can observe that there is a symmetry, in which the results of $\langle n^\prime\rangle = 3/8$ with $h_0 < 0$ map onto the ones with $\langle n^\prime\rangle = 5/8$ with $h_0 >0$, and is readily observed via a particle-hole transformation $\hat b_{i}^{\phantom{\dagger}} \to \hat b_{i}^\dagger$ in the Hamiltonian~\eqref{eq:ham}. 

At small $V/J$, the absence of $\pi$-phase shift is likely expected since the interactions are not sufficiently large to induce a strong (even if short-ranged)  density wave pattern. Furthermore, the imposition of a stripe ($h_0\neq 0$) is seen to be detrimental to pairing, where the zero momentum occupancy decreases from its maximum value at $h_0 = 0$. At strong interactions $V/J = 4$, on the other hand, charge density-wave formation is robust, and depending on the strength of the stripe chemical potential, a $\pi$-phase shift is observed at sufficiently positive (negative) $h_0$ for total density $\langle \hat n^\prime\rangle = 3/8$ ($\langle \hat n^\prime\rangle = 5/8$). This is seen via the concomitant reversal in sign of the correlations $c_{\bf i_0}(2,0)$ and $c_{\bf i_0}(2,1)$ at a critical value $h_0^c$ of the stripe chemical potential: $h_0^c \simeq 1.3J$ for $\langle \hat n^\prime\rangle = 3/8$ and $h_0^c \simeq -1.3J$ for $\langle \hat n^\prime\rangle = 5/8$. That is, $c_{\bf i_0}(2,0)$ turns negative, whereas $c_{\bf i_0}(2,1)$ becomes positive, denoting the phase flip of the charge ordering in the interstripe regions.

Once an anti-phase checkerboard solid phase sets in [shaded regions in Fig.~\ref{fig:ED_result_V1_4}(e--h)], a substantial reduction in the zero momentum occupancy occurs, pointing out to an apparent competition of $\pi$-phase shift and superfluid behavior. This agrees with the results in the main text, which showed that also with \textit{spontaneously} formed stripes, stripes with phase-reversal solids are incompatible with a superfluid. 

Another important aspect is whether the onset of $\pi$-phase shift constitutes a phase transition. In the SSE results on the spontaneous formation of stripes, we mainly investigate densities where discontinuities on the $\langle \hat n\rangle$ vs.~$\mu$ curves emerge~\cite{Scalettar1995, Sengupta2005}, that were interpreted as signaling a first-order phase transition between a superfluid and solid phases~\cite{Hebert2001, Kohno1997, Schmid2002, Kuklov2004}. In other lattice geometries, such transitions were later identified with first~\cite{Wessel2005} or weak first-order~\cite{Isakov2006} for triangular and kagome lattices, respectively. Here, when the stripes are `externally' imposed, one can directly check the nature of the transition to a $\pi$-phase solid as the stripe chemical potential $h_0$ is varied. 

\begin{figure}[t] 
   \includegraphics[width=1.0\columnwidth]{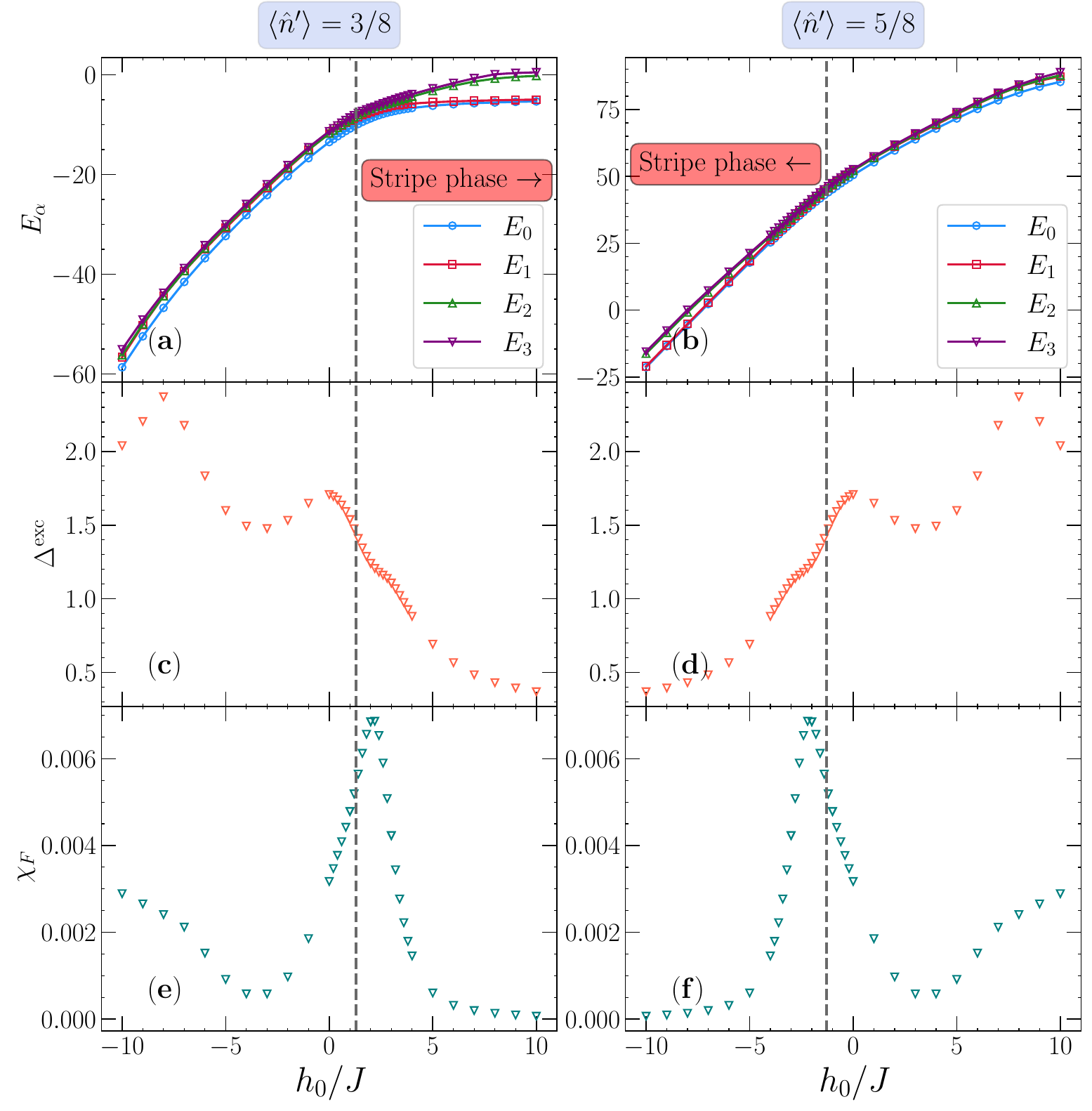}
   \caption{(a) [(b)] The low-lying eigenspectrum $E_\alpha$ in the $\textbf{q}=(0,0)$ quasi-momentum sector of the Hamiltonian for $\langle \hat n^\prime\rangle = 3/8$ [$\langle \hat n^\prime\rangle = 5/8$] filling as a function of the stripe onsite energy $h_0$; (c)[(d)] the excitation gap $\Delta^{\rm exc}$ which is finite across a wide range of $h_0$. (e) and (f) give the fidelity susceptibility for the two densities investigated. The dashed vertical lines mark the regime where the $\pi$-phase shift sets in, given by the sign reversal of $c_{\bf i_0}(2,0)$ and $c_{\bf i_0}(2,1)$ correlations -- see Fig.~\ref{fig:ED_result_V1_4}.}
   \label{fig:ED_spectra_gaps_fid}
\end{figure}

Figure~\ref{fig:ED_spectra_gaps_fid} displays how the low-lying eigenspectrum $E_\alpha$ at the zero-quasi momentum  $\textbf{q}=(0,0)$ sector of the Hamiltonian (wherein the ground state resides) depends on $h_0$ for the two fillings investigated. No signatures of level crossing emerge, thus ruling out a first-order phase transition at this lattice size, as can be seen by the finiteness of the excitation gap $\Delta^{\rm exc} \equiv E_1 - E_0$ across a wide range of stripe chemical potential values but especially at the critical point $h_0^c$ where the sign-reversal of the $c_{i_0}(2,0)$ and $c_{i_0}(2,1)$ correlations sets in. Lastly, we can investigate how the many-body ground-state $|\Psi_0\rangle$ evolves upon variations of $h_0$. This is computed by the fidelity susceptibility~\cite{Zanardi06, CamposVenuti07, Zanardi07, You2007},
\begin{equation}
\chi_{F} = \frac{2}{L_xL_y} \frac{1 - |\langle \Psi_0(h_0)|\Psi_0(h_0+dh_0)\rangle|}{dh_0^2},
\end{equation}
with $dh_0 = 10^{-3}J$ in our calculations; it determines a continuous quantum phase transition location (without any information about the type of symmetry-breaking) through the locus of an extensive peak in the system size in the region parameters of interest~\cite{Yang07, Varney2010, Jia11, Mondaini2015, Jin2022}. In the case of first-order phase transitions, it exhibits discontinuities that are proportional to $dh_0$. Here, with a single system size available, we can see that the $\chi_F$ peak location is not directly at the point of the emergence of $\pi$-phase shift, but rather can be traced to the inflection point of the zero momentum occupancy $n_{{\bf k}=0}$ in Figs.~\ref{fig:ED_result_V1_4}(f) and (h). It signals that a potential (second-order) phase transition is associated with the suppression of superfluidity instead.

\subsection{SSE results}
\begin{figure}[t] 

   \includegraphics[width=1.0\columnwidth]{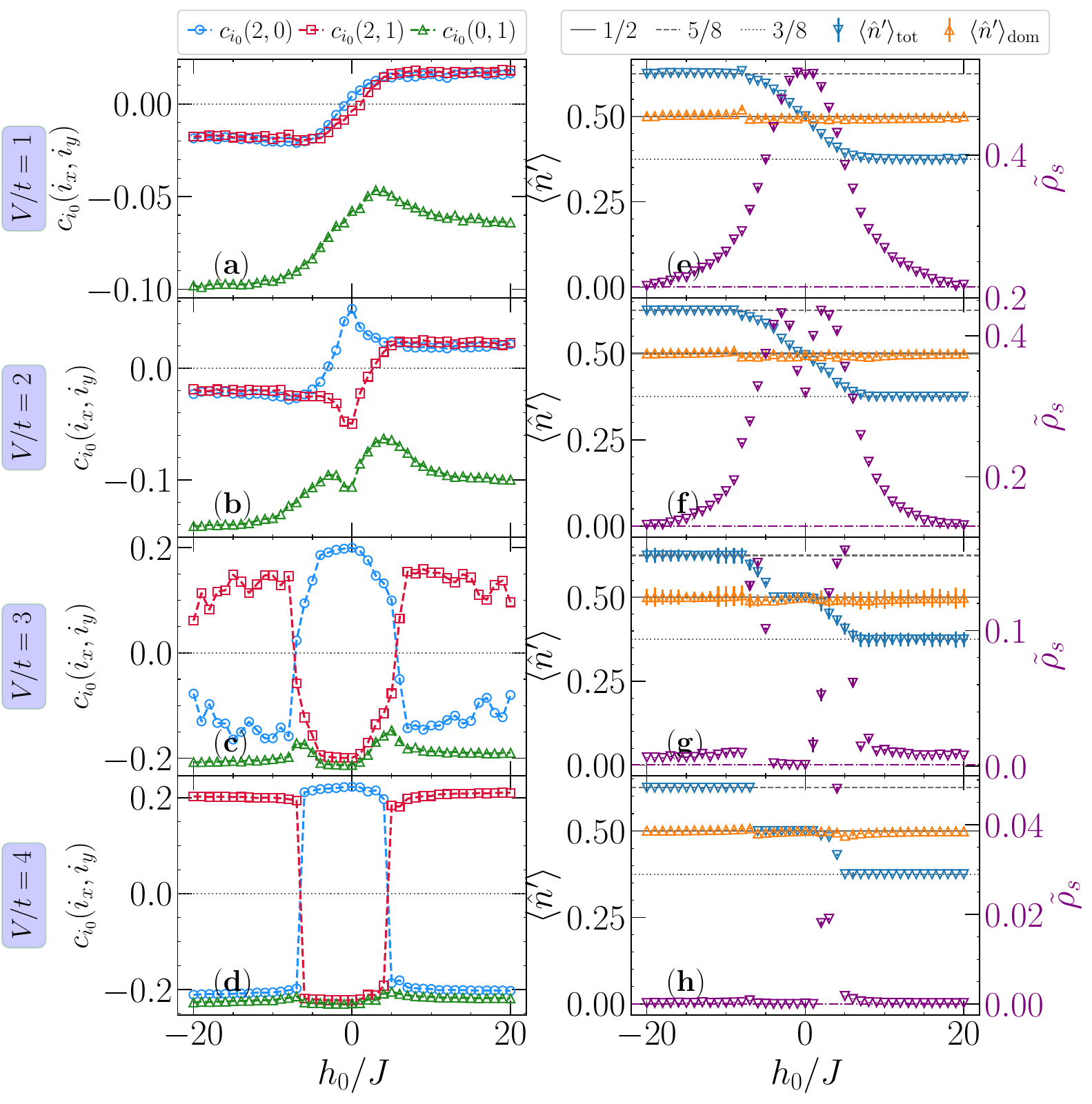}
   \caption{Results from the SSE calculations on an $8 \times 8$ lattice. Left panels (a--d) show the correlations, similar to the ones in the ED results, vs.~the stripe chemical potential $h_0$ for increasing repulsive interactions $V/J = 1, 2, 3$ and 4. The right panels  (e--h) show the total density $\langle \hat n^\prime\rangle$ and the inter-stripe density $\langle \hat n^\prime\rangle_{\rm dom} \simeq 1/2$ for the corresponding interaction strengths of the left panels. The right axes signal the superfluid density $\tilde \rho_s$.}
   \label{fig:SSE_result}
\end{figure} 

While the system sizes amenable to the ED calculations are limited, here we tackle a full two-dimensional case $8\times 8$ using SSE. As previously pointed out, these simulations have a slightly different approach: We set a global chemical potential to obtain an interstripe density $\langle \hat n^\prime\rangle_{\rm dom} = 1/2$, as shown in the right panels of Fig.~\ref{fig:SSE_result}, for whichever value of the set of parameters $(V, h_0)$. In this case, the $\pi$-phase shift also takes place, mainly when the total density is $\langle \hat n^\prime\rangle = 5/8$ and $h_0 <0$ or when $\langle \hat n^\prime\rangle = 3/8$ and $h_0 >0$, within the regime of strong interactions ($V/J \gtrsim 2$). This thus mimics the ED results at the same total density. Note, however, that taking $|h_0|\to\infty$ would lead to phase separation, and the correlations across a stripe would eventually converge to zero.

Similar to the ED results, once the $\pi$-phase shift sets in, the superfluid weight is substantially suppressed [see Figs.~\ref{fig:SSE_result}(g) and \ref{fig:SSE_result}(h), right vertical axes]. The only regimes where it becomes finite are when the total density is transitioning among the different plateaus, analogously to what happens via changing the global chemical potential in the absence of added stripes~\cite{Sengupta2005}. These results confirm that `stripe crystals' (static stripes with phase flip) cannot generally be reconciled with superfluidity for hardcore boson models.

\section{$s$-wave pairing in the fermionic case}
\begin{figure}[t] 
\vskip0.2in
   \includegraphics[width=1.0\columnwidth]{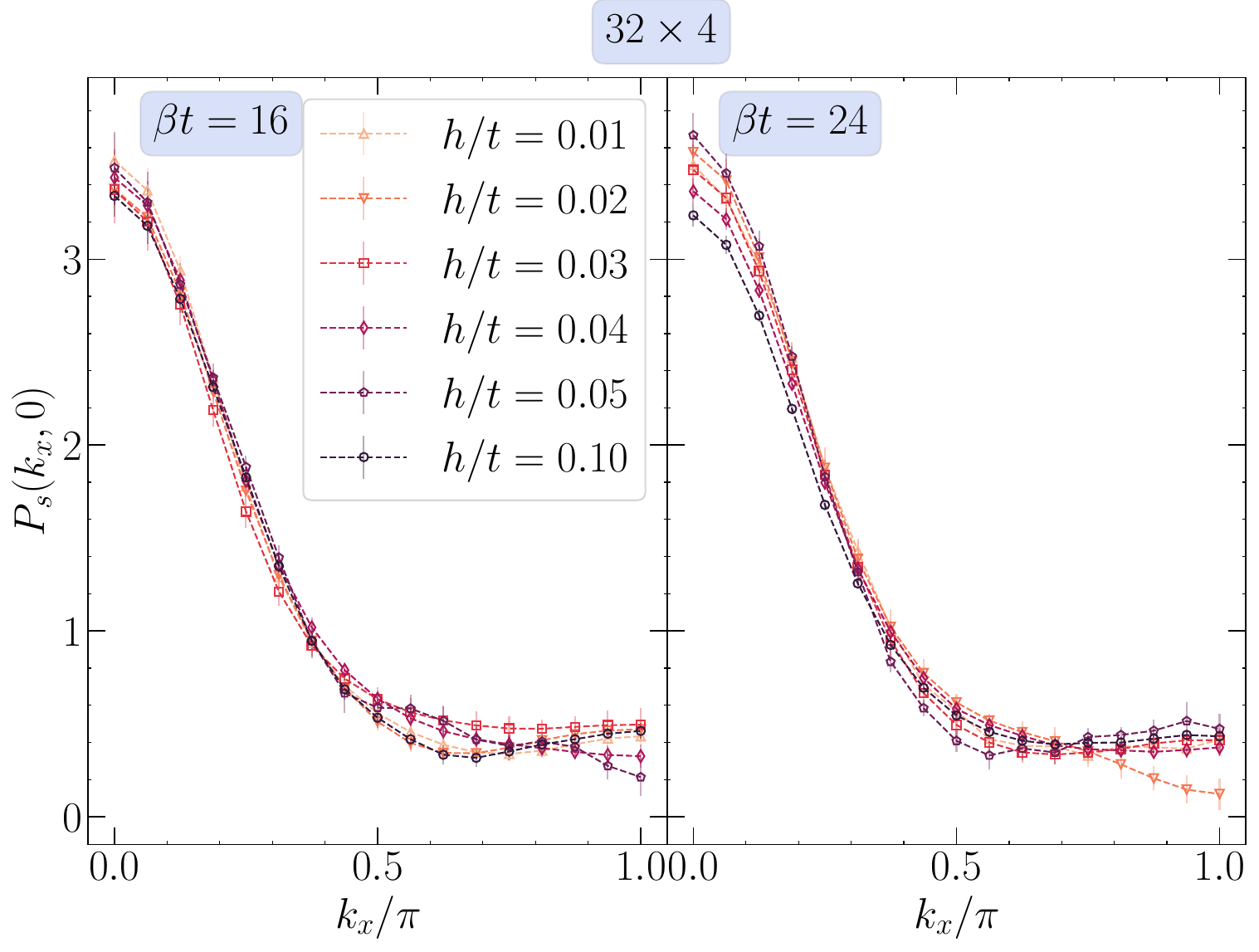}
   \caption{Pair structure factor of the fermionic model at $k_y = 0$ vs.~$k_x$ for (left) $\beta t = 16$ and (right) $\beta t = 24$. Different magnitudes of the antiphase pinning field $h$ (see main text for definition) are contrasted. Data are obtained at $|U|/t = 15$, with imaginary-time discretization $t\Delta t = 0.05$, and hole-doping $\delta = 1/8$ on a $32\times 4$ ladder with periodic boundary conditions.}
   \label{fig:P_s_finite_momentum}
\end{figure}

While many similarities emerge between the results of actual cuprates, such as $\pi$-phase shift in local orders, and their interplay with pairing, one aspect of the current fermionic model studied, which contrasts with them, concerns the modulation of pairing. In cuprates, there exists an indication toward a pair-density-wave modulation with the same period of the spin stripes~\cite{Tranquada2020}. If that were the case also in the attractive Hubbard model, one would see a peak of the pair structure factor $P({\bf k}) = \sum_{i, j} e^{{\rm i} {\bf k}({\bf r}_i-{\bf r}_j)}\langle  \hat\Delta_i^{\phantom{\dagger}} \hat \Delta_j^\dagger\rangle/(L_xL_y)$ at finite-momentum values. As indicated in Fig.~\ref{fig:P_s_finite_momentum}, one does not observe such a situation: the peaks are mainly seen at $k_x = k_y = 0$, mapping instead to the macroscopic occupancy of the zero-momentum modes for the case of a Bose-Einstein condensation.

\end{document}